\documentclass[12pt,a4paper,onecolumn,notitlepage,oneside]{article}
\input tcilatex

\begin{document}

\title{A Modified Theory Of Newtonian Mechanics}
\author{Amir H. Abbassi\thanks{%
E-mail: ahabbasi@net1cs.modares.ac.ir}$\;\;\;\&\;$ Amir M. Abbassi\thanks{%
Permanent add: Depatment of Physics, Faculty of Sciences, Tehran University}$%
\;^,\;$\thanks{%
E-mail: amabasi@khayam.ut.ac.ir} \\
\textit{Department of Physics, School of Sciences, }\\
\textit{Tarbiat Modarres University,P.O.Box 14155-4838}\\
\textit{Tehran, Iran}}
\date{}
\maketitle

\begin{abstract}
 A specific model of the inertial law is presented by which we can have
some deeper insight into the essence of mass and inertia.
In this modified theory there is no need to keep the concept of
absolute space and the third law as a principle. By introducing a 
convenient form for gravitational law the coupling constant G
becomes a function of inertial parameters of the universe.  \\

\end{abstract}


According to the  Newton's second law of motion the relation 
$\vec F =m\vec a$ holds between the applied force on a particle and its
acceleration in absolute space and any other frames which are related to
it by Galilean transformations. These reference frames are called inertal 
frames and $m$ is the inertial mass of the particle. The relation between 
coordinates of two inertial systems $S$ and $S^{\prime}$ which are moving 
with constant velocity $\vec v$ with respect to each other are :

\begin{equation}
\left\{
\begin{array}{l}
t^{\prime}=t  \\
\vec a^{\prime}=\vec a  \\
\vec u^{\prime}=\vec u -\vec v  \\
\vec x^{\prime}=\vec x -\vec v t +\vec x_0
\end{array}
\right.
\end{equation} \label{1}

Despite  its simple appearance and practical applications Newton's 
concept of absolute space which is the basis of Newtonian mechanics
(NM) and in turn the basis of mass and energy is not well defined and has 
been criticized. Cogent arguments against absolute space are \cite{1}:
\begin{itemize}
\item[a]- There is no unique way of locating Newton's absolute space
within the finite class of inertial frames.
\item[b]- It conflicts with one's understanding to conceive of a thing
which acts but cannot be acted upon.
\end{itemize}

In his critique of Newton's conceptions in {\it``The Science of Mechanics"},
Mach suggested a set of pioneering implicit ideas then addressed by A. Einstein
as {\it Mach's Principle} (MP) \cite{2}. According to MP, every motion is only comprehensive as a relative motion, inertia is due to an interaction with
average mass of the
universe \cite{3} and the distribution of matter in the universe determines the inertial frame at each point. There have been many attempts to find Machianized reformulations of mechanics\cite{4,5,6}. Einstein's main aim in the  general theory of relativity (GR) was to do this through giving an equation for
gravitation and inertia together. But it came out that his theory shows 
some non-Machian aspects and do not fulfill the MP. For instance it has
solution for empty space.  Also there are many alternative theories
of gravitation motivated by Mach's ideas \cite{7,8,9}. Since in GR as our standard
theory of gravitation we are still faced with unresolved questions such as
singularity problems and lack of well defined source for inertia, then it is
reasonable to have more discussions about foundations. It seems with closer 
scrutiny of Newton's second law we may find some insight.  

Actually in this work our aim is to present a simple method for Machianization 
of NM. We propose that inertial effect is a mutual 
interaction between two particles which in any non-rotating arbitrary
reference frame $S$ is proportional to the difference between their
accelerations  and to the inertial charges of each
individual particles as follows
\begin{equation}
\vec F_{inertia}=\mu . c_1 . c_2 (\vec a_1 - \vec a_2)
\end{equation}  \label{2}
\noindent where $\vec a_1$ , $\vec a_2$ are accelerations of particles 1 and 2
with respect to $S$ , their inertial charges are denoted by $c_1$ , $ c_2$
respectively , and $\mu$  is an inertial coupling constant. 
This can be easily extended to systems consisting of N particles.
Again in any non-rotating arbitrary reference frame $S$ we have
\begin{equation}
\vec F_i =\mu . c_i \sum_{j=1}^{N} c_j (\vec a_i - \vec a_j)
\end {equation}  \label{3}
\noindent where $\vec F_i$ is the applied force on particle i and the summation 
is done over all particles.

In the real world the inertial charge and Newtonian inertial mass of a particle
 are related so that
\begin{equation}
m_i =\mu. c_i \sum_{j=1}^{all}c_j 
\end{equation} \label{4}
where summation is done over all particles in the universe. Since local inhomogenities have no observed effects on the inertial mass
then it is accepted that the inertial mass is determined by the global 
structure of the universe and the relation (4) for the inertial mass is in 
accordance with this general belief.  This can be accounted as  a simple formulation
of Mach's idea about inertia.  In terms of inertial mass , equation (3) can be
rewritten in the following form
\begin{equation}
 \vec F_i = m_i \left(\vec a_i-\frac {\sum\limits_{j=1}^{all}m_j \vec a_j}{
\sum\limits_{j=1}^{all}m_j}\right)
\end{equation}  \label{5}
These new forms of the second law  i.e. equations (3) and (5) are 
invariant under transformation to a more general group of reference frames  $S^{\prime}$ 
than Galilean ones which we may call as generalized Galilean transformations.
\begin{equation}
\left\{
\begin{array}{l}
t^{\prime}=t \\
\vec a^{\prime}=\vec a - \vec b \\
\vec u^{\prime}=\vec u - \vec b t -\vec v \\
\vec x^{\prime}= \vec x -\frac12\vec b t^2 -\vec v t +\vec x_0
\end{array}
\right.
\end{equation} \label{6}
Here $\vec b$ , $\vec v$ and $\vec x_0$ are the constant acceleration ,velocity
and position of $S^{\prime}$ with respect to $S$ at $t=0$ respectively.

Within the set of reference frames $S^{\prime}$ with different values of 
$\vec b$ and $\vec v$ there is a unique subset call it $S^{\prime\prime}$
moving with acceleration $\vec b_0$
\begin{equation}
\vec b_0 = \frac{\sum\limits_{j=1}^{all} m_j \vec a_j}{\sum\limits_{j=1}^{all} m_j}
\end{equation}  \label{7}
with respect to $S$. In $S^{\prime\prime}$ the form of Newton's second 
law of inertia is recovered, $\vec F_i =m \vec a^{\prime\prime}$. 
Therefore, inertial frames are a unique set of frames which are moving
with acceleration $\vec b_0$ with respect to $S$ with $\vec v$ and $\vec x_0$
be any arbitrary value. This again may be accounted for as  a  feasible
formulation of Mach's idea about inertial frames. Since $\vec b_0$ is 
determined globally then local inhomogeneities have a negligible 
effect on it and inertial frames.

It is evident that equations (3) and (5) also satisfy Newton's third law.
For a two particle system we have
\begin{equation}
\vec F_1 = -\vec F_2 =\mu . c_1 . c_2 (\vec a_1 - \vec a_2)
\end{equation}  \label{8}
and for a system with N particles it makes
\begin{equation}
\sum_{j=1}^{all}\vec F_i =0
\end{equation}  \label{9}
So there is no need to introduce Newton's third law as an extra principle.

We can extend this model to Newton's law of gravitation. 
Equivalence principle here means that the source of inertia and 
gravitation is the same. Then if we define the gravitational force 
between two particles of inertial charges $c_1$ and $c_2$ as
\begin{equation}
\mid\vec F_G\mid= \frac{\mu^2 . c_1 . c_2}{\mid\vec r_{12}\mid^2}
\end{equation}  \label{10}
where $\mid\vec r_{12}\mid=\mid\vec r_1 -\vec r_2\mid$ is the relative 
separation between particles 1 and 2.  Then we can express gravitational
constant $G$ in terms of inertial charges $c_i$s
\begin{equation}
G=\left(\sum_{j=1}^{all} c_j\right)^{-\frac12}
\end{equation}
which may be accounted for as another aspect of Machian implications
that the so-called physical constants of nature (like G) are to be 
determined with some global features of the universe. Since the inertial mass and the gravitational constant
are finite quantities, this means that $\left(\sum\limits_{j=1}^{all} c_j\right)$ is finite too,
and the universe cannot be infinitely extended.

The Lagrangean formalism  based on this model of inertial
law is the same as the one for NM except that the kinematic energy of system
T which is equal to $\sum\limits_{i} \frac12 m_i {v_i} ^2$ is replaced by
\begin{eqnarray}
T&=&{\sum\limits_i \frac12 m_i {v_i}^2}-\frac{[\sum\limits_i m_i {\vec v_i}]^2}
{2\sum\limits_i m_i}  \\
&=&\frac14\sum\limits_i \sum\limits_j m_i m_j \frac{(\vec v_i -\vec v_j)^2}
{\sum\limits_j m_j}  \nonumber
\end{eqnarray}
This is an invariant scalar. Then  total energy of the system i.e. the sum of 
kinetic energy T and potential energy $V(r_{ij})$ , a function of relative
separation $r_{ij} = \vec r_i -\vec r_j$, is invariant in all non-rotating 
reference frames.

\begin{center}
\bf{Remarks}
\end{center}
\begin{itemize}
\item[(i)]- It may seem that choosing non-rotating frames is some kind of
restriction which reduces the generality of the chosen frames. This is 
however, not the case, since rotating frames with respect to the whole
universe are distinguishable and can be fixed.
\item[(ii)]- Our results show that $G$ and $m$ are not constants of nature but 
may change anytime the total inertial charge of the universe changes significantly.
This can happen in the early stage of evolution of the universe, i.e. in the epochs of 
pair production.    
\item[(iii)]- The concept of energy introduced with this theory is
independent of the reference frame.
\item[(iv)]- There is no inertial strucure for empty space.

\end{itemize}

These may show us how some thing should change in a modified theory 
of relativity.  

\end{document}